%% file: main.tex
\documentclass[runningheads]{llncs}
\usepackage[T1]{fontenc}

\usepackage{graphicx}
\usepackage{xcolor}
\usepackage{cite}
\usepackage{hyperref}

\usepackage{todonotes}

\input{preamble}%
\begin{document}

\newcommand{\juffte}{juFFTe}
%
\title{Performance Evaluation of Fast Fourier Transforms on Emerging RISC-V Hardware with Vector Extension Support}
%
%
\author{%
    Daniel Seibel\orcidID{0000-0003-3833-2089} \and
    Kaveh Haghighi Mood\orcidID{0000-0002-8578-4961} 
    Jayesh Badwaik\orcidID{0000-0002-5252-8179} \and
    Prateek Chawla\orcidID{0000-0001-8895-2791} \and
    Stepan Nassyr\orcidID{0000-0002-0035-244X} \and
    Andreas Herten\orcidID{0000-0002-7150-2505}%
}
\authorrunning{D. Seibel et al.}
%
\institute{J{\"u}lich Supercomputing Centre, J{\"u}lich, Germany
\email{\{k.haghighi.mood,d.seibel\}@fz-juelich.de}}
\maketitle              

\begin{abstract}
This manuscript presents a performance evaluation of Fast Fourier Transform (FFT) implementations on emerging processors supporting the RISC-V Vector Extension (RVV~1.0). 
By introducing \juffte, a light-weight high-performance library for discrete Fourier transforms, it is demonstrated how effective vectorization of performance-critical FFT kernels can be achieved on RVV-enabled hardware.
Comprehensive benchmarks on three RVV~1.0-ready processors, the SiFive X280, the X100 core of the SpacemiT K3 and the C920v2 core of the Sophon SG2044, reveal substantial performance improvements of \juffte{}\footnote{\url{https://github.com/FZJ-JSC/juFFTe}} over the widely used FFTW3 library.
Although RVV-enabled platforms show promising results at this stage of development, a comparison with AMD's Zen 5 architecture indicates that RISC-V needs further maturing to reach the performance of established micro-architectures.
\end{abstract}
\section{Introduction}
The discrete Fourier transform (DFT) is one of the most fundamental mathematical operations in natural sciences and serves as the backbone of many numerical algorithms thanks to its efficient implementation in the fast Fourier transform (FFT)~\cite{VanLoan1992, Takahashi2019}.
FFTs play a crucial role in modern high performance scientific computing landscape, including the solution of partial differential equations found in computational mechanics~\cite{Kabel2014}, electromagnetics~\cite{Seibel2022}, computational chemistry~\cite{petersen1995}, and physics~\cite{abraham2011,Giannozzi2017,Buchheit2023}. 
Being a fundamental building block of modern HPC software stacks, it is widely recognized as one of the HPC dwarfs~\cite{colella2004,Asanovic2006} and the availability of highly-optimized FFT libraries is often considered a key indicator of the maturity of a software ecosystem targeting high performance.
Besides scientific and engineering applications, the FFT is at the center of digital signal processing~\cite{Proakis1996} and image processing~\cite{Reddy1996}.

Owing to its central role in numerous applications and libraries, optimizing FFT requires careful consideration of both algorithmic design and hardware-specific enhancements to ensure consistently high performance across platforms.
In the FFT framework, large DFTs are split into smaller DFTs, which constitute the computationally intensive part of the FFT algorithm and, hence, must be carefully optimized for the target hardware. 
On modern CPUs and accelerators, this typically entails efficient vectorization using intrinsics or assembly tailored to the specific architecture.

RISC-V is a relatively new instruction set architecture released under a permissive license that allows for royalty-free hardware development.
Although RISC-V has been primarily used in low-power devices, there are ongoing efforts to develop a new generation of chips targeted at the high-performance segment~\cite{DARE, EUPILOT}.
To this end, the RISC-V Vector extension (RVV)~\cite{rvv} has been ratified in 2021 and added to the RVA23 profile~\cite{rva23} to enable widespread support for  vectorization on future RISC-V hardware.
Notable features of RVV include the concept of vector-length-agnostic code and the support of variable vector lengths, which avoids the difficulties associated with different vector lengths encountered in vector extensions like AVX and AVX512~\cite{intel_sdm}.
With regards to FFT, RVV offers special movement and permutation instructions that have the potential to significantly speed up the complex multiplications and transpositions part of the performance-critical kernels.

Despite the prospect of high-performance RISC-V hardware on the horizon, currently available production FFT libraries provide only limited support for the RISC-V ecosystem. 
At the time of writing, only the open-source library FFTW3~\cite{FFTW2005} has been ported. 
This manuscript addresses this gap by introducing \juffte{}, a modern Fortran library inspired by FFTE subroutines~\cite{takahashiffte, TakahashiFranchetti2020}. \juffte{} reimplements several methods described in FFTE, using modern Fortran (Fortran 90 and later), improving maintainability and usability while preserving the performance-oriented design of the original implementation. 
With regards to RISC-V, \juffte{} leverages the RISC-V Vector Extension in version 1.0 (RVV~1.0)~\cite{rvv} to enable the efficient and performant vectorization of DFT kernels.
As part to its modular structure, these performance-critical kernels are isolated from the remaining code, making the library as portable as possible while preserving a high level of performance and optimization.
Moreover, it features a unified API and supports the popular FFTW interface to serve as a drop-in replacement.

In addition to the introduction of \juffte{}, the manuscript's main contribution is the evaluation of the proposed implementation on state-of-the-art RISC-V hardware, namely the SiFive X280, the SpacemiT K3 and the Sophon SG2044, all supporting RVV~1.0.
In various experiments, the performance of \juffte{} is analyzed in single and multi-core scenarios and compared against those of FFTW, offering new insights about the hardware as well as vectorization with RVV.
It is demonstrated that \juffte{} can yield average performance improvements of up to a factor of three depending on hardware due to superior vectorization and parallelism.
To the best knowledge of the authors, this represents the first in depth study of high-performance FFT on RVV~1.0 capable hardware.

The paper is structured as follows: After a brief overview of related work on FFT for RVV and a short introduction to the theory of FFT, the optimizations carried out in \juffte{} for RVV are reviewed with a special focus on vectorization.
This is followed by a series of benchmarks, where the performance of the library is compared against the FFTW library across different RVV-enabled systems and an x86 reference system.


\section{Related Work}
FFT has been evaluated on multiple RISC-V platforms using both the base ISA and architectural extensions. Jiang et al.~\cite{Jiang2023} proposed an ISA extension consisting of twelve additional instructions to accelerate FFT on the open-source RISC-V NutShell processor~\cite{NutShell}. In a separate work, Jiang et al.~\cite{Jiang2025} introduced Zoozve, an extension to RVV, demonstrating significant improvements in FFT performance. With the emergence of RVV, several studies have explored vectorized FFT implementations. Vizcaino et al.~\cite{vizcaino2023} investigated different Radix-2 FFT approaches on an RVV-ready prototype and compared performance with the NEC SX-Aurora VE vector architecture. Perotti et al.~\cite{Perotti2024b} evaluated FFT benchmarks on Ara2, an open-source processor compliant with RVV~1.0. At the library level, Zhao et al.~\cite{Zhao2023} optimized PerfMPL-FFT for power-of-two transform sizes on the C910MP CPU~\cite{c910} and compared performance against the widely used FFTW library~\cite{FFTW2005}. Cross-platform comparisons have also been conducted to evaluate the competitiveness of emerging many-core RISC-V processors. Strack et al.~\cite{Strack2026} compared FFTW's performance on conventional x86 CPUs and the Sophon SG2042 processor. Similarly, Brown ~\cite{Brown2025} evaluated the FFT component of NASA’s Parallel Benchmark suite~\cite{NASP2009} and compared the performance of the Sophon SG2044 and SG2042 with x86 and ARM processors.

We further advance these efforts by showcasing the \juffte{} library with support for RVV~1.0 to bring high-performance FFT support to RISC-V and by evaluating performance across multiple novel RVV-ready hardware implementations with a systematic comparison against FFTW.

\section{Fast Fourier Transform}
The DFT of a complex vector
$x = {(x_0,\ldots, x_{N-1})}^\top\in\mathbb{C}^N$ is defined by
\[
  y_k = \sum_{j=0}^{N-1} \omega_N^{kj} x_j, \quad k=0,\ldots,N-1,
\]
where the complex coefficients $\omega_N^{kj} = \exp \left(-2\pi \imath\, kj / N\right)$ with $\imath^2 = -1$ are called twiddle factors.
It can be written as a matrix-vector product $y = F_N x$ with $F_N[p,q] = {\left(\omega_N^{pq}\right)}$.
If $N = r c$, then $x_{r\times c}[i,j] = x_{rj + i}$
is obtained by reshaping $x$ into a column-major-ordered $r\times c$
two-dimensional array.
The central idea underpinning all FFT algorithms is the so-called radix splitting~\cite{VanLoan1992, Takahashi2019}.
If $N = r m$ and $x\in\mathbb{C}^N$, then the DFT of length $N$ can be split into two DFTs of length $r$ and $m$ as follows
\begin{equation}\label{eq:mixed_radix}
  {\left( F_N x\right)}_{m\times r} = \left[ F_N[0:m-1,0:r-1]
    \odot \left(F_m x_{r\times m}^\top \right)\right] F_r.
\end{equation}
Here, $\odot$ denotes the Hadamard product or point-wise multiplication of two arrays.
The recursive application of \autoref{eq:mixed_radix} leads to the mixed-radix splitting algorithms and variations like the four-step and six-step algorithms~\cite{VanLoan1992, Takahashi2019}.

	



There is an inherent degree of freedom in how the kernels access and store the intermediate data and this aspect leads to different FFT frameworks.
Besides the widely known Cooley-Tukey algorithm~\cite{Cooley1965}, the Stockham algorithms~\cite{Stockham1966} have been particularly popular for vector hardware, because they access memory contiguously and avoid expensive data rearrangement like bit-reversal permutations.
This comes at the cost of a larger memory footprint as an additional workspace
vector is required.
Nonetheless, most numerical libraries feature auto-sorting kernels.
Among open-source efforts, FFTE~\cite{TakahashiFranchetti2020} uses the
decimation-in-frequency Stockham algorithm to target SIMD-enabled hardware and
VkFFT~\cite{Tolmachev2023} implements the transposed Stockham algorithm for a wide-range of hardware.
Exemplary C code for a radix-$2$ Stockham kernel with pre-computed twiddle factors is given in~\autoref{lst:stockham_serial}.
It can be seen from the code that the computation is divided into two stages, the multiplication with the so called butterfly matrix $F_2$ and the point-wise multiplication with the twiddle factors.
Since the kernels encapsulate the computational part of the FFT, it is of critical importance to optimize them to improve performance.

\begin{lstlisting}[float,style=cstyle,caption={Radix-$2$ Stockham kernel},label={lst:stockham_serial}]
void /*FUNC*/stockham2/*ENDFUNC*/(int n, int l, int r, double complex *x, double complex *y, double complex *tw) {
    int lp = l / 2, rp = r * 2;
    for (int j = 0; j < lp; j++) {
        // Twiddle factors come in pairs, but the first is always 1
        const double complex tw0 = 1., tw1 = tw[j];
        for (int k = 0; k < r; k++) {
            double complex x0 = x[j * r + k], x1 = x[(lp + j) * r + k];
            // Apply F_2 butterfly
            double complex z0 = x0 + x1, z1 = x0 - x1;
            // Multiply twiddle with factors
            y[j * rp + k] = tw0 * z0;
            y[j * rp + r + k] = tw1 * z1;
        } 
    }
}
\end{lstlisting}


\section{Library Optimizations}
\juffte{} can leverage vectorization in two complementary ways. 
First, it benefits from compiler-assisted auto-vectorization, allowing modern compilers to generate vector instructions directly from high-level Fortran code. 
Second, it can employ vectorized kernels written in C or even assembly code with explicit use of vector instructions to further enable fine-grained control over vector operations and efficient utilization of hardware capabilities.
Together, these approaches allow \juffte{} to effectively exploit vector units and achieve improved performance portability across different platforms.

Since writing manually vectorized kernels is cumbersome and error-prone, the kernels for RISC-V have been generated with the help of {SPIRAL}~\cite{Spiral}, a domain-specific language and framework for the automatic generation for highly optimized and vectorized DFT kernels.
In order to support the RISC-V vector extension, SPIRAL has been extended with C-intrinsics for RVV~1.0 based on the work for ARM`s vector extension SVE~\cite{TakahashiFranchetti2020}.
The generated kernels are vector-length agnostic, which means that they are binary compatible with every implementation of RVV~1.0 regardless of chosen vector length.
One particularly useful feature of RVV are the so-called segmented load and store instructions, which operate directly on interleaved complex floating‑point data while avoiding additional shuffle instructions. 

\begin{lstlisting}[float,style=cstyle,caption={Serial and vectorized packing (block-transposition) kernels.},label={lst:autovec}]
// Extract columns i to i + bx from x and store as rows in z
void /*FUNC*/pack/*ENDFUNC*/(int i, int bx, int ny, int ld, const double complex *x, double complex *z) {
    // x and z are column-major 
    for (int k = 0; k < bx; k++)
        for (int j = 0; j < ny; j++)
            z[k * ld + j] = x[j * nx + i + k];
}

void /*FUNC*/v_pack/*ENDFUNC*/(int i, int bx, int ny, int ld, const double complex *x, double complex *z) {
	for (int j = 0; j < ny; j += vl) {
		size_t vl = /*FUNC*/__riscv_vsetvl_e64m1/*ENDFUNC*/(ny - j);
		for (int k = 0; k < bx; k++) {
			vfloat64m1x2_t xv = /*FUNC*/__riscv_vlsseg2e64_v_f64m1x2/*ENDFUNC*/(&x[j * nx + i + k], 16 * nx, vl);
			/*FUNC*/__riscv_vsseg2e64/*ENDFUNC*/(&z[k * ld + j], xv, vl);
		}
	}
}
\end{lstlisting}

In addition to the kernels generated with SPIRAL, explicit use of C-intrinsics for RVV was required in several situations specific to RVV.
In contrast to the handling of AVX-512, both GCC/GFortran (ver. 15.2.0) and Flang (ver 20.1.8) are not able to effectively vectorize block-transposition operations with complex floating-point data like in~\autoref{lst:autovec}. 
One possible explanation is that segmented load and store operations are treated as expensive in the compiler’s vectorization cost models.
Instead of issuing strided vector loads and stores, they use regular unit-strided loads and stores on single \texttt{complex double} elements at a time.
Since this essentially restricts the vector length to 128 bits, the full potential of RVV is not realized, making the manual use of RVV intrinsics necessary to achieve effective vectorization.
The routine \texttt{v\_pack()} in~\autoref{lst:autovec} shows how the available vector length can be fully utilized with strided segment loads (\texttt{vlsseg}) and unit-stride segment stores (\texttt{vsseg}) in a vector-length-agnostic manner.

It should be noted that future compiler releases are likely to solve this issue by adapting internal cost models that disfavor strided segment loads and stores.
Nonetheless, \juffte{} allows developers to perform these kinds of necessary optimizations thanks to its modular structure, which isolates performance critical kernels.


\section{Benchmarks}
The main performance indicator of an FFT implementation is the time $t$ it takes to compute the DFT depending on the transform size.
Since the space dimension plays a minor role in the context of vectorization, the standard complex-to-complex DFT transform in 1D with varying size $N$ is selected as the benchmark problem.
The corresponding performance in FLOP/s can be approximated by $5 N \log_2(N) / t$, which is in fact an upper bound.
In order to obtain consistent measurements, the time necessary to compute a 1D DFT of size $N$ is recorded across multiple runs and averaged afterwards.
The performance of \juffte{} is evaluated against an optimized version of FFTW3 with RVV~1.0 support based on the work by R.~Dolbau~\cite{dolbeaufftw3}.
All computations are carried out in double precision, with multi-threading enabled via OpenMP for both libraries.
Except where explicitly stated otherwise, FFTW uses the default \texttt{FFTW\_MEASURE} planner flag.


    




%
For the evaluation of \juffte, three different RVV~1.0 capable CPUs are considered, the SiFive X280 cores available on the Tenstorrent Blackhole devices, the X100 high-performance cores of the SpacemiT K3 processor and the C920v2 cores of the Sophon SG2044 processor.
Their features and capabilities are summarized in~\autoref{tab:cpus}, with the peak compute and memory bandwidth measured experimentally.

\begin{table}[htb]
\scriptsize
\centering
\caption{Overview of the RISC-V CPUs under investigation. Bandwidth is bi-directional (read + write).}
\begin{tabularx}{\linewidth}{>{\raggedright\arraybackslash}X *{3}{>{\centering\arraybackslash}X}}
    \toprule%
    & {\textbf{SiFive X280}} & {\textbf{SpacemiT K3}} & {\textbf{Sophon SG2044}} \\
    \midrule%
    Core & SiFive X280 & X100 & C920v2 \\
    \# Cores & 4 & 8 & 64 \\
    Frequency & 1.75 GHz & 2.4 GHz & 2.6 GHz \\
    VLEN (DLEN) & 512 (256) bit & 256 (128) bit & 128 (128) bit \\
    Peak FP64 (per core) & 13.98 GLFOP/s & 17.58 GLFOP/s & 20.07 GLFOP/s\\
    L1 size & 32 kiB (per core) & 64 kiB (per core) & 64 kiB (per core) \\
    L1 bandwidth (per core) & 55.90 GB/s & 44.51 GB/s & 35.66 GB/s \\
    L2 size & 128 kiB (per core) & 4 MiB (per 4 cores) & 2 MiB (per 2 cores) \\
    L2 bandwidth (per core) & 55.82 GB/s & 28.83 GB/s & 28 GB/s \\
    L3 size & 2 MiB (per 4 cores) & n/a & 64 MiB (per 64 cores) \\
    L3 bandwidth (per core) & 25.00 GB/s & n/a & 5.6 GB/s \\    
    \bottomrule%
\end{tabularx}\label{tab:cpus}
\end{table}

\subsubsection{SiFive X280:}
The first system under consideration consists of the RISC-V margin nodes of the Tenstorrent Blackhole, which houses four tiles of four SiFive X280 cores each, for a total of 16 cores.
The SiFive X280 core is a 8-stage dual-issue in-order 64-bit RISC-V CPU with a vector length (VLEN) of 512 bit and 256 bit datapath (DLEN).
The vector engine contains one arithmetic unit capable of four FP64 FMAs per cycle and a separate vector load and store unit. 
The resulting peak compute and memory throughput are listed in~\autoref{tab:cpus}.
Each core has 32 kiB of L1 data and 128 kiB of L2 cache, both of which operate with the same bandwidth of 256 bits per cycle or 55.9 GB/s.
The L3 cache with a size of 2 MiB is shared between the four cores in a tile with a bandwidth of 25 GB/s per core.

\begin{figure}[htb]
\centering%
\includegraphics[width=0.6\textwidth]{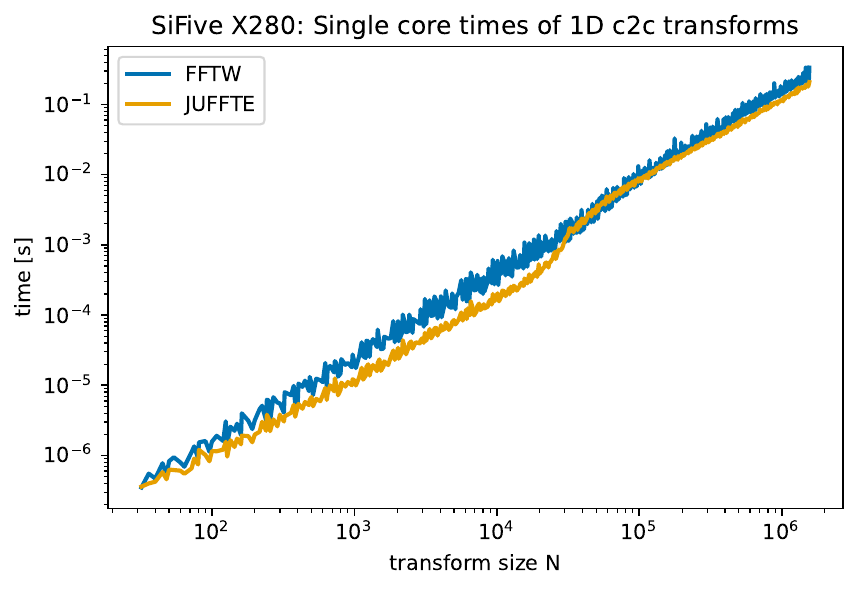}

\centering%
\includegraphics[width=0.6\textwidth]{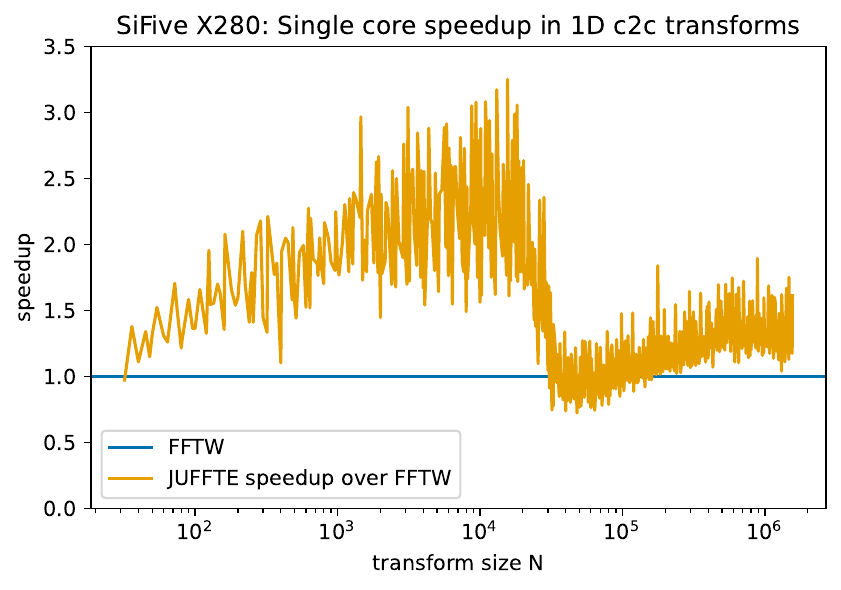}
\caption{Execution times (top) and speedup (bottom) of {\juffte} over FFTW on SiFive X280 with a single thread.}\label{fig:sifive_single}
\end{figure}

\begin{figure}[htb]
\centering%
\includegraphics[width=0.6\textwidth]{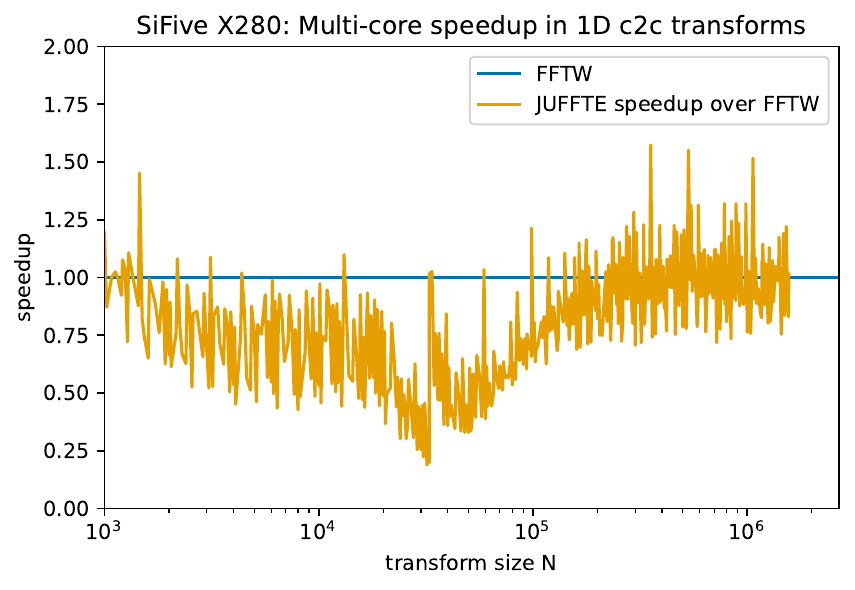}
\caption{Speedup of \juffte{} over FFTW on SiFive X280 with four threads.}\label{fig:sifive_omp_speedup}
\end{figure}

In \autoref{fig:sifive_single}, the absolute time in seconds to compute a 1D DFT of given size $N$ on a single core is plotted in a log-log scaling in the top plot.
The relative speedup per transform size is shown in the bottom figure with logarithmic scaling for the $x$-axis, with values greater than one indicating that \juffte{} is faster than FFTW for given $N$.
It can be seen that \juffte{} outperforms FFTW consistently up to a size of $N\approx 16000$, which corresponds to the limit of the L2 cache of 128 kiB as a complex double precision vector of length $N$ takes $16N$ Bytes of memory.
The reason for the superior performance of \juffte{} over FFTW lies in the use of segmented vector load and store instructions, which perform particularly well on the SiFive X280 and yield more bandwidth than unit-stride memory operations.
Once the bounds of the L2 cache are passed, \juffte{} and FFTW exhibit similar levels of performance.
As the memory latency increases significantly for accesses outside the L3 cache, the computations become memory-bound for large $N$ and most time is spent waiting for data to arrive from and to DRAM.
Despite lacking dedicated prefetching capabilities, the SiFive X280 benefits from packing routines part of the six-step algorithm implemented in \juffte{}, which allow the computational kernels to stay in the L2 cache.
As a result, \juffte{} gains the lead for larger $N$ as the additional costs of the packing procedures pay off.
By enabling multi-threading with OpenMP, \juffte{} and FFTW show similar levels of performance for larger $N$ as can be seen from \autoref{fig:sifive_omp_speedup}, indicating that \juffte{} could benefit from FFTW's greater variety of parallelized FFT algorithms that are selected during its auto-tuning stage. 
Note that at the time of this writing, only four out of the 16 X280 cores present on the Blackhole device are accessible due to software limitations.

\subsubsection{SpacemiT K3 -- X100:}
The SpacemiT K3 is a multi-core 64-bit RISC-V CPU featuring two types of cores, the X100 and the A100~\cite{spacemitk3}.
The X100 is a four-issue out-of-order 64-bit RISC-V CPU based on the OpenC910 with a 256 bit vector length (VLEN) and 128 bit datapath length (DLEN).
With two arithmetic units and two load/store units, the peak performance is four FP64 FMAs per cycle, see also~\autoref{tab:cpus}.
The L1 data cache of each X100 is 64 kiB big and four X100 share 4 MiB of L2 cache, which is the last level of cache.
There are eight X100 and eight A100 cores present on the K3, making a total of 16 cores.
Since early testing revealed inferior performance for double precision computations compared to the X100 cores, the results for the A100 are omitted here.
The K3 is not commercially available at the time of writing and the data presented here has been gathered on a cloud platform as part of the official beta test program.


\begin{figure}[htb]
\centering%
\includegraphics[width=0.6\textwidth]{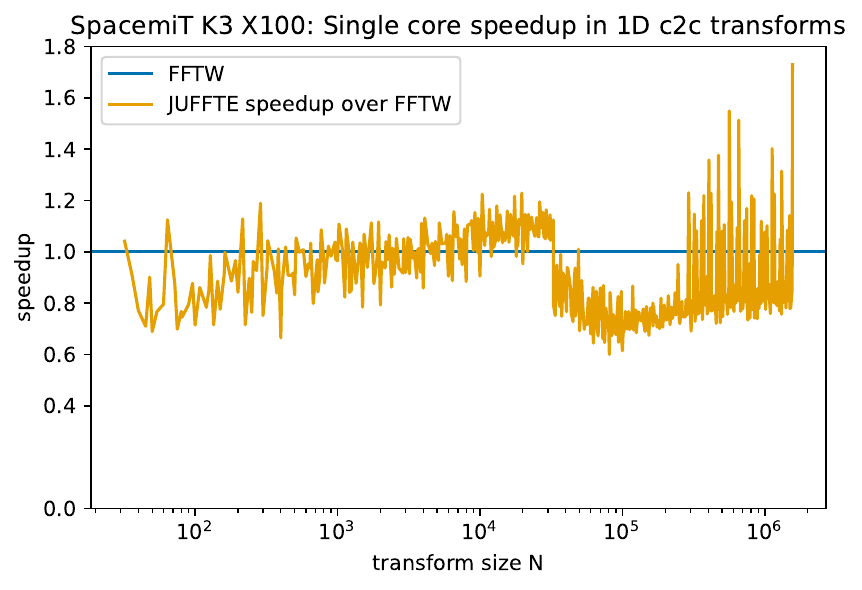}
\caption{Speedup of \juffte{} over FFTW on SpacemiT X100 with a single thread.}\label{fig:k3_single_speedup}
\end{figure}

\begin{figure}[htb]
\centering%
\includegraphics[width=0.6\textwidth]{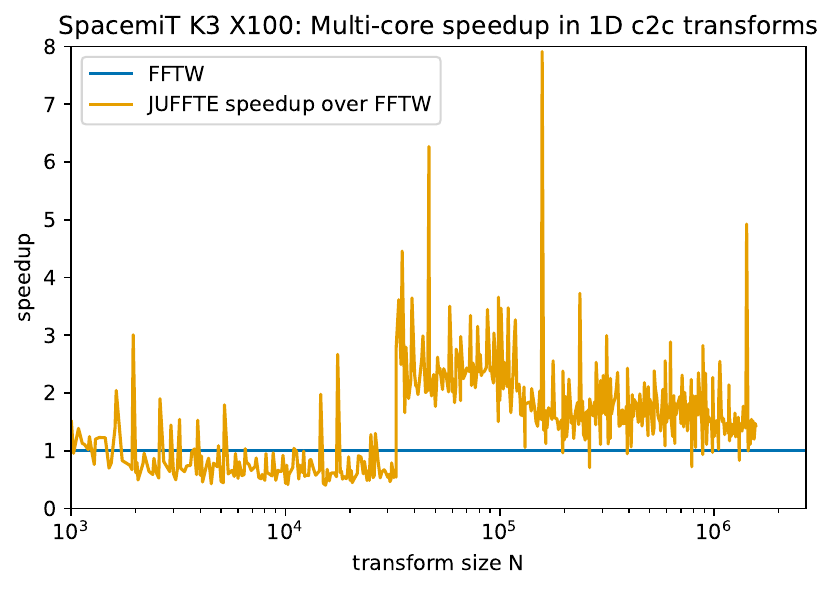}
\caption{Speedup of \juffte{} over FFTW on SpacemiT X100 with eight threads.}\label{fig:k3_omp_speedup}
\end{figure}

The relative speedup on a single core of \juffte{} over FFTW for the SpacemiT X100 is visualized in \autoref{fig:k3_single_speedup} with logarithmic scaling of the $x$-axis.
Although FFTW is faster for very small $N$, \juffte{} is more performant for sizes that almost fill up the L1 cache.
In contrast to the SiFive X280, the advantage of segmented loads and stores disappears on the SpacemiT X100 and FFTW's use of unit-stride memory instructions combined with dedicated in-lane shuffling for complex computations remains competitive.
Within the limits of the L2 cache, FFTW has the upper hand as it deploys kernels of larger radix while \juffte{} is limited to radices less than or equal to 16.
The advantage of larger radices lies in the reduced number of total memory operations, which is critical as FFTs are usually memory bound.
Although larger radices are technically feasible for \juffte{}, the use of segmented instructions effectively doubles the number of vector registers in use, causing register spillage as only 32 vector registers are available for RVV.
For these reasons, it follows that \juffte{} should incorporate the conventional vectorization strategy of using unit-stride loads and stores with shuffling instructions as an alternative in the future.
For large $N$, the picture is mixed with \juffte{} or FFTW being up to 80\% faster depending on the particular $N$.
In \autoref{fig:k3_omp_speedup}, respective results for the multi-core case with eight cores are presented.
Due to limited access time on the test system and the extensive overhead of FFTW's auto-tuning with multi-threading, only heuristic planning has been invoked with \texttt{FFTW\_ESTIMATE}.
Therefore, \juffte{} has a substantial lead over FFTW for larger $N$ in L2 being almost 50\% faster in most cases.

\subsubsection{Sophon SG2044:}
The Sophon SG2044 is a recent processor targeting high-performance computing applications. 
It is equipped with 64 Xuantie C920v2 cores, which are 12-stage out-of-order multiple issue superscalar 64-bit RISC-V CPUs~\cite{Brown2025}. 
In contrast to its predecessor, the SG2042, the SG2044 fully supports RVV~1.0 with a vector length of 128 bits (VLEN).
The peak floating point performance of four FP64 FMAs per cycle indicates that two vector arithmetic units are present on each core.
There are also two vector load/store units.
Each core is equipped with 64 kiB of L1 data cache and four cores share 2 MiB of L2 cache.
All cores share 64 MiB of L3 cache.
The reported bandwidths in~\autoref{tab:cpus} are much lower than for the other two CPUs tested and are yet only obtainable for certain specific message sizes.
This issue needs to be investigated further.
At the time of submission, the SG2044 is the only commercially available server-grade RISC-V processor implementing the RVV~1.0 vector extension.
All experiments and performance evaluations were conducted on the Monte Cimone cluster~\cite{Venieri2026} using Sophon nodes equipped with the SG2044 processor.

\begin{figure}[htb]
\centering%
\includegraphics[width=0.6\textwidth]{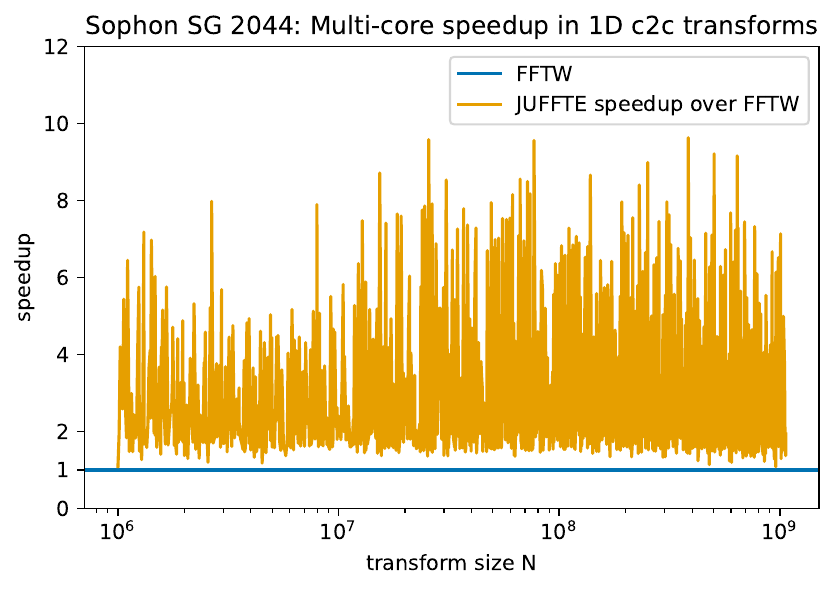}
\caption{Speedup of \juffte{} over FFTW on Sophon SG2044 with 64 threads.}\label{fig:sophon_omp_speedup}
\end{figure}

Since the Sophon SG2044 features a high core count of 64 cores with 128 GB of DRAM, only the multi-core results are considered here, while single-core results are summarized in \autoref{tab:speedup}.
\autoref{fig:sophon_omp_speedup} shows the relative speedup of \juffte{} over FFTW against the transform size $N$ in logarithmic scaling for the $x$-axis.
Again due to limitations of FFTW's auto-tuning, the setting has been set to \texttt{FFTW\_ESTIMATE} to invoke heuristic planning.
For every $N$ benchmarked in the range from $10^6$ to more than $10^9$, \juffte{} computes DFTs faster than FFTW with an average speedup of 3x.
This suggests that the six-step algorithm in \juffte{} is especially effective for higher core counts in a shared-memory environment.
Also, the use of packing and unpacking to keep DFT computations in L2 cache amortizes for large $N$ as the number of data accesses per element increases with the transform size $N$.

\subsubsection{Comparison against Zen 5:}
In order to understand how well these RVV~1.0 capable CPUs perform, a comparison against AMD`s Zen 5 architecture is made.
The Strix Point Zen 5 cores found in the HX 370 CPU are 8-wide out-of-order x64 CPUs that support the AVX-512 SIMD extension with a 256 bit datapath.
The theoretical peak performance is about 8 FP64 FMAs per cycle or 82.5 GFLOP/s per core with a maximum core frequency of 5.16 GHz~\cite{zen5}.
The cache bandwidth is 32 Bytes per cycle per core just as the SiFive X280, but due to the higher clock rate, the theoretical maximum bandwidth of the L1 (48 kiB) and L2 (1 MiB) cache is around 163 GB/s per core.
The four Zen 5 cores in the HX 370 share 16 MiB of L3 cache with about half the theoretical bandwidth of the L1/L2 caches. 
Since the Zen 5 core can clock at much higher frequencies than the RISC-V CPUs, the maximum frequency is capped to 2.6 GHz to give a more indicative comparison.
Nonetheless it should be noted that the HX370 is a mobile CPU with a similar TDP as the SiFive X280 and SpacemiT K3.

\begin{figure}[htb]
\centering%
\includegraphics[width=0.6\textwidth]{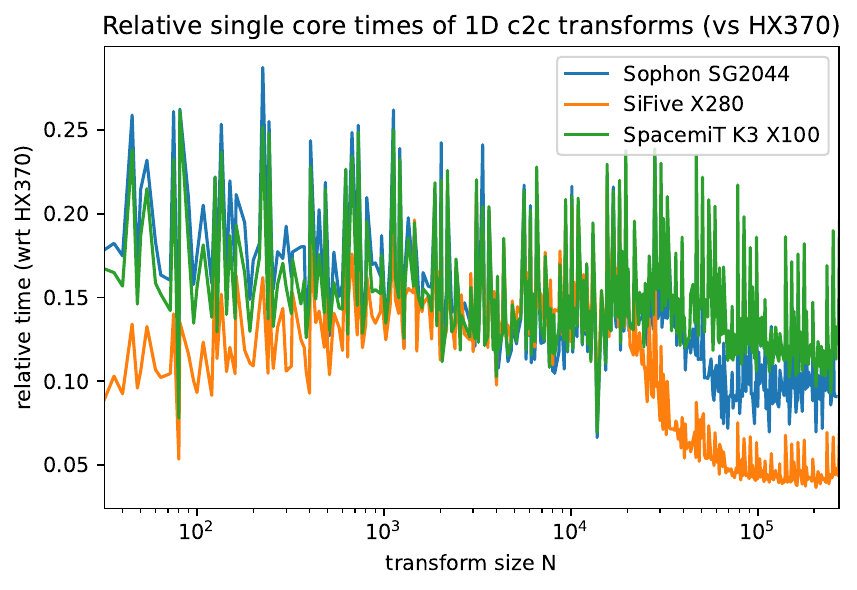}
\caption{Relative execution times $t_{\textup{Zen 5}} /  t_{\textup{RISC-V}}$ with a single thread.}\label{fig:comparison}
\end{figure}

\autoref{fig:comparison} compares the single-core performance of the RISC-V CPUs with the Zen 5 cores of the HX370.
For each system, the plotted curve shows the point-wise minimum of the time-to-solution of \juffte{} and FFTW on the Zen 5 core relative to RISC-V for given transform sizes $N$, i.e. $t_{\textup{Zen 5}} /  t_{\textup{RISC-V}}$, again with logarithmic scaling for the $x$-axis.
This ensures that the best possible FFT is chosen for every $N$.
While the RISC-V CPUs all show a similar level of performance, the Zen 5 core is significantly faster throughout the entire range of transform sizes.
Reasons for this are manifold as the Zen 5 cores have double the floating-point throughput of 16 FLOP per cycle, a higher cache bandwidth and a much more intricate front-end with powerful reordering capabilities and branch prediction.
Nevertheless, the C920v2 core of the Sophon SG2044 and the SpacemiT X100 manage to attain about 15\% of the performance of the Zen 5 core in average.
The SiFive X280 core has difficulties keeping up for small and large $N$ which can mostly be attributed to its comparatively low frequency and low DRAM bandwidth.
These results demonstrate that the RISC-V ecosystem for high-performance computing still has room to catch up to established platforms like x86, but the emergence of RVV~1.0-capable hardware is an important step in the right direction, see also~\cite{Brown2025}.

\section{Conclusion}
As shown in \autoref{tab:speedup}, \juffte{} can achieve considerable speedups over FFTW in single- and multi-core DFTs for RVV~1.0-enabled RISC-V CPUs.
However, the performance evaluation also demonstrates that the optimization of FFT algorithms for RVV needs to take different vectorization strategies into account.
RVV offers several ways of vectorizing DFT kernels due to its variety of special strided memory operations and permutation instructions, which perform differently depending on the hardware.
A thorough analysis based on advanced roofline models is therefore needed to identify performance bottlenecks and adapt the vectorization to the capabilities of the hardware. 
For \juffte{}, the authors aim to implement various vectorization strategies for RVV in combination with an auto-tuner similar to FFTW in order to attain high-performance for existing and future RISC-V hardware.
Since auto-vectorization offered by compilers has shown to be unreliable in certain situations, the authors also plan to create a generator for DFT kernels in assembly with explicit RVV instructions to ensure maximum performance.

\begin{table}[htb]
\centering
\caption{Best average speedup (single or multi-core) of \juffte{} over FFTW.}\label{tab:speedup}
\begin{tabularx}{\textwidth}{l *{3}{>{\centering\arraybackslash}X}}
\toprule
 & SiFive X280 & SpacemiT K3 X100 & Sophon SG2044 \\
\midrule
Single-core avg.~speedup & 1.53x & 0.90x & 0.73x \\
Multi-core avg.~speedup & 1.09x & 1.91x & 3.03x \\
\bottomrule
\end{tabularx}
\end{table}

\begin{credits}
\subsubsection{\ackname} 
The authors gratefully acknowledge the Monte Cimone project for providing access to the Sophon compute nodes used in this work. 
They thank SpacemiT and the RISC-V Ecological Application Innovation Center in Zhujiang for early access to the SpacemiT K3 via the remote beta test program.
Funding for parts of this work has been received from EuroHPC’s project DARE
SGA 1 under Grant Agreement No.~101202459.

\subsubsection{\discintname}

\end{credits}
%
%
%
\bibliographystyle{splncs04}
\bibliography{mybibliography}
%
\end{document}

%% file: preamble.tex
\usepackage[english]{babel}
\usepackage{mathtools}
\usepackage{amsfonts}
\usepackage{listings}
\definecolor{cbblue}{RGB}{0,102,204}   
\definecolor{cborange}{RGB}{230,97,0}  
\definecolor{cbcyan}{RGB}{0,153,153}   
\definecolor{cbgreen}{RGB}{0,153,51}   
\definecolor{bg}{RGB}{247,247,247}
\lstdefinelanguage{CColorBlind}{
  language=C,
  morekeywords={size_t,vfloat64m1x2_t},
  sensitive=true
}
\lstdefinestyle{cstyle}{
  backgroundcolor=\color{bg},
  basicstyle=\fontsize{7pt}{8.4pt}\ttfamily\selectfont,
  numbers=left,
  numberstyle=\tiny\color{gray},
  numbersep=6pt,
  frame=single,
  rulecolor=\color{black!30},
  frameround=tttt,
  language=CColorBlind,
  keywordstyle=\color{cbblue}\bfseries,
  commentstyle=\color{cborange}\itshape,
  stringstyle=\color{cbcyan},
  identifierstyle=\color{black},
  moredelim=[is][\color{cbgreen}]{/*FUNC*/}{/*ENDFUNC*/},
  showstringspaces=false,
  breaklines=true,
  tabsize=3,
  captionpos=b
}

\usepackage{tabularx}
\usepackage{booktabs}